# Integrating Atomic Layer Deposition and Ultra-High Vacuum Physical Vapor Deposition for *In Situ* Fabrication of Tunnel Junctions

Alan J. Elliot[1,a], Gary A. Malek[1], Rongtao Lu[1], Siyuan Han[1], Haifeng Yiu[2], Shiping Zhao[2], and Judy Z. Wu[1,b]

[1] Department of Physics and Astronomy, The University of Kansas, Lawrence, KS 66045

[2] Beijing National Laboratory for Condensed Matter Physics, Institute of Physics, Chinese Academy of Sciences, Beijing 100190, China

Atomic Layer Deposition (ALD) is a promising technique for growing ultrathin, pristine dielectrics on metal substrates, which is essential to many electronic devices. Tunnel junctions are an excellent example which require a leak-free, ultrathin dielectric tunnel barrier of typical thickness around 1 nm between two metal electrodes. A challenge in the development of ultrathin dielectric tunnel barrier using ALD is controlling the nucleation of dielectrics on metals with minimal formation of native oxides at the metal surface for high-quality interfaces between the tunnel barrier and metal electrodes. This poses a critical need for integrating ALD with ultra-high vacuum (UHV) physical vapor deposition. In order to address these challenges, a viscous-flow ALD chamber was designed and interfaced to an UHV magnetron sputtering chamber via a load lock. A sample transportation system was implemented for *in situ* sample transfer between the ALD, load lock, and sputtering chambers. Using this integrated ALD-UHV sputtering system, superconductor-insulator-superconductor (SIS) $Nb/Al/Al_2O_3/Nb$ Josephson tunnel junctions were fabricated with tunnel barriers of thickness varied from sub-nm to ~ 1 nm. The suitability of using an Al wetting layer for initiation of the ALD $Al_2O_3$ tunnel barrier was investigated with ellipsometry, atomic force microscopy, and electrical transport measurements. With optimized processing conditions, leak-free SIS tunnel junctions were obtained, demonstrating the viability of this integrated ALD-UHV sputtering system for the fabrication of tunnel junctions and devices comprised of metal-dielectric-metal multilayers.

**I. INTRODUCTION**

Many technologies, both mature and nascent, rely on ultrathin (~ 1 nm) dielectric layers to act as tunnel barriers between two electrodes to form metal-insulator-metal (MIM) structures. For example, magnetic tunnel junctions (MTJs), which are wholly responsible for the rapid miniaturization of computer memories, are simply two metallic ferromagnetic thin film electrodes with a ~1-2 nm dielectric layer between them [1]. The figure-of-merit tunnel magnetoresistance (TMR), defined as the ratio of the resistance of the device when the ferromagnetic layers are magnetized in parallel and anti-parallel directions, depends critically on the thickness of the dielectric layer. The TMR oscillates with the thickness of the dielectric layer with a period of only ~0.3 nm [2], so subnanometer thickness control of ultrathin films is necessary. Another example is the Josephson junction (JJ), a superconductor-insulator-superconductor (SIS) device used in voltage standards, superconducting quantum interference devices (SQUIDs), and, most recently, quantum bits (qubits) [3]. A leak-free tunnel barrier with thickness much smaller than the superconducting coherence length is typically required for the superconductor electrodes to remain phase coherent. Further, because, the critical current through the JJ decays exponentially with increasing tunnel barrier thickness [4], in Nb-Al/AlOx/Nb JJs the AlOx tunnel barrier thickness is typically on the order of 1 nm [5].

---

[a] Corresponding author: Alan J. Elliot. Electronic mail: alane@ku.edu
[b] Corresponding author: Judy Z. Wu. Electronic mail: jwu@ku.edu

Producing an ultrathin, uniform, and leak-free dielectric film is difficult on metal substrates due to the naturally formed native oxides on most metals such as Nb. Nb-Al/AlOx/Nb JJs are an excellent example. FIG 1a depicts schematically a Nb-Al/AlOx/Nb JJ. In order to form AlOx tunnel barrier, a few nanometers of Al is sputtered *in situ* on the Nb bottom electrode to serve as a wetting layer, and AlOx is formed by exposing this wetting layer to a controlled pressure of $O_2$ *in vacuo*. This thermal oxidation scheme has been used to create high quality JJs using either Nb or Al as electrodes. These JJs have been the building blocks for a large variety of commercialized superconducting devices. SQUIDs represent one of these successes and have been used widely for detection of extremely small magnetic signals [6]. When such JJs are employed for qubits, a more stringent requirement for lower noise arises to avoid superconducting phase decoherence. One major source of noise is oxygen vacancies in the AlOx tunnel barrier, which are formed by oxygen diffusion during the thermal oxidation process. These vacancies act as two-level-fluctuators and catastrophically couple the qubit to the environment [7], destroying the entanglement on which quantum computation relies and drastically increasing the computational error rate. In order to improve the JJ-based qubits, an alternative fabrication scheme must be adopted to generate a defect free, uniform and ultrathin tunneling barrier (FIG 1b).

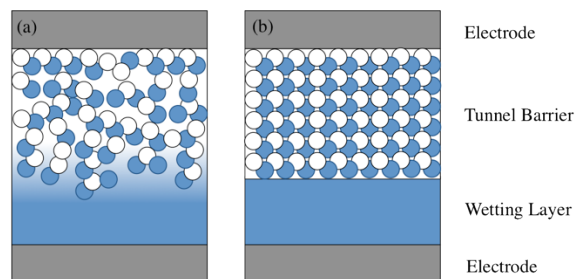

**FIG 1** In traditional Josephson junction (JJ) fabrication techniques, an Al wetting layer is exposed to oxygen to produce a tunneling barrier of aluminum oxide. This produces an inhomogeneous film (a) with oxygen vacancies and interstitials that lead to decoherence in JJ qubits. A uniform film (b) will reduce the density of these defects and produce a more coherent qubit. New fabrication techniques need to be explored to produce such a tunnel barrier.

There are several alternative schemes for fabricating high-quality MIM trilayer stacks, one of which is Molecular Beam Epitaxy (MBE). MBE relies on the very tightly controlled sublimation of solid sources in an ultra-high vacuum, allowing atomic layer-by-layer heteroepitaxy of different materials in the stack. Luscher reviewed the basic considerations of MBE chamber design in 1979 [8] and much work has been done since then, including the design of *in situ* substrate exchangers for multiple sample fabrication [9] and implementation of characterization tools such as scanning electron microscopy for *in situ* microstructure characterization [10]. MBE has been applied to many materials including III-V [11], and II-VI semiconductors [12], as well as complex high temperature superconductors like Yttrium Barium Carbon Oxide (YBCO) [13]. While MBE can be used to grow MIM structures, it is a remarkably expensive process, which limits MBE's applicability in small scale research and high-end electronics commercialization. Chemical Vapor Deposition (CVD) (see a comprehensive review in 2003 by Choy [14]) has been widely used in coating of functional materials in the single layer or multilayer films. CVD works by exposing a sample in a low vacuum chamber to a gaseous flow of sources which react at the substrate surface. CVD can create dense, pure materials with high growth rates and uniformity and is capable of growing many different materials including metals (Cu, Al, etc.), dielectrics ($Al_2O_3$, $SiO_2$, etc.), semiconductors (Si, GaN, etc) and even superconductors such as TiN. [14]. CVD growth of multilayer stacks, including MIM, SIS and even metal-insulator-semiconductor, is commonly reported.[15] However, it is difficult to control the growth rate of CVD to achieve subnanometer precision in layer thickness. In the context of CVD growth, "ultrathin" is usually defined as <10 nm, with several reports claiming ultrathin films of thicknesses as low as 4 nm [16]. In its current form, CVD is therefore unsuitable for the fabrication of the ultrathin tunnel barriers required for MTJs and JJs.

Atomic Layer Deposition (ALD) is also a chemical process, but it differs from CVD in terms of its self-limiting growth mechanism, which allows thickness precision at the atomic scale. ALD produces atomic layer-by-layer growth via sequential exposure of relevant chemical sources following well-defined chemical reactions. Taking $Al_2O_3$ as an example, alternating pulses of $H_2O$ and trimethylaluminum (TMA) are exposed to heated substrates, separated by a flush of inert carrier gas to assure the two chemicals never meet in a gaseous state. Growth of $Al_2O_3$ occurs via ligand exchange between $H_2O$ and TMA at the sample surface and is described by the chemical reactions[17]

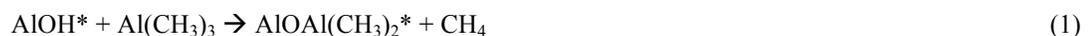  (1)

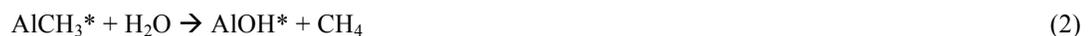  (2)

$$AlOH^* + Al(CH_3)_3 \rightarrow AlOAl(CH_3)_2^* + CH_4 \quad (1)$$

$$AlCH_3^* + H_2O \rightarrow AlOH^* + CH_4 \quad (2)$$

where an asterisk denotes a surface species. There are several unique merits associated with the ALD process. First, ALD is a relatively low temperature process with ALD $Al_2O_3$ typically occurring near 200 °C. This low thermal budget is particularly important to monolithic devices on Si-readout circuits. Another merit is that the involved chemical reactions occur only on the sample's surface, and the reactions stop at the completion of each exposure. ALD growth is hence self-limiting. In each cycle of ALD $Al_2O_3$, *i.e.* after both the reactions shown in Equations (1) and (2) have occurred once, only one molecular layer is produced, or about 1.2 Å thickness of $Al_2O_3$. This provides atomic-scale control of film thickness. Finally, ALD coatings are highly conformal, which is particularly important to coating surfaces with large aspect ratios. A large variety of films, including metals and dielectrics, can be grown using ALD as long as the sources for the relevant chemical reactions are available [18]. ALD is therefore an excellent candidate for growth of truly ultrathin insulating layers as tunnel barriers of atomic resolution and has been reviewed in depth multiple times [17,18].

The quality of ultrathin films depends critically on their nucleation on substrates (or "M" electrode in MIM structures), which means substrate surface preparation is a key towards achieving leak-free tunnel barriers using ALD. The chemical reactions in an ALD process (for example, ALD $Al_2O_3$ given in eqns. 1-2) require the existence of surface species, particularly hydroxyl surface groups (OH*) or methyl surface groups ($CH_3$*). This requirement is automatically satisfied on certain substrates, such as $SiO_2$ since residual $H_2O$ on the surface produces a well hydroxylated surface ready for ALD nucleation. However, for substrates that are poorly hydroxylated, such as hydrogen terminated silicon (H-Si), nucleation is frustrated due to the lack of reaction sites on the surface. While the dangling hydrogen bonds on H-Si do serve as reaction sites to some degree, the initial stages of growth are dominated by the formation of a ~1 nm thick silicate interfacial layer (IL)[19]. However, surface activation, such as pre-exposing the H-Si to a large dose of TMA for ALD $Al_2O_3$ growth, has been shown to reduce the IL to ~0.5 nm for a $Al_2O_3$ film with a total thickness of ~3 nm [19,20]. Similarly to $SiO_2$ and H-Si, metallic substrates can be classified into two categories; those with a reactive surface, such as Al and Cu, and those without, such as Au and Pt. In the former case, for *ex situ* deposited metals, a native oxide of several nanometers (up to ~5 nm for Al) will pre-exist, and ALD growth occurs easily on top [21,22]. For *in situ* deposited metals, an IL may form from thermal oxidation or chemisorption of the ALD precursors, and this IL may range in thickness from ~0.4 nm on *in situ* ALD-W [23] to ~2 nm on *in situ* sputtered Al [22]. On noble metals, such as Pt, Ir, and Ru, nucleation of ALD films can be completely frustrated during the first 30-50 cycles of growth [24]. These initial cycles act as an incubation process to prepare the surface for nucleation by adsorbing source material on the surface, effectively increasing its reactivity. ILs several nm thick are commonly reported when growing ALD dielectric films on noble metals, and they form through the diffusion of source material into the metal film, such as the diffusion of Tetrakis(ethylmethylamido)hafnium(IV) (TEMAH) into Pt during the growth of $HfO_2$. [25,26]. The exact thickness and composition of the IL depend on the substrates and sources used. But, in the case of ALD-$HfO_2$ on Pt, the IL thickness can be reduced from ~10 nm to ~5 nm and the interface can be made more uniform by exposing the metal film to a hydrous plasma to promote surface oxidation before ALD dielectric layer growth.[26]. In either case of reactive or noble metals, the IL issue must be addressed in order to produce an ultrathin dielectric tunnel barrier using ALD on a metal substrate with minimized IL effect for tunnel junctions and many other MIM structures.

Considering the difficulties in surface activation of noble metals for nucleation of ALD, a very thin wetting layer of non-noble metal with controllable formation of IL layers seems a possible resolution to obtain MIM tunnel junctions with ultrathin ALD dielectric tunnel barriers if the "M" layer itself is either a noble metal or has complicated surface chemistry leading uncontrollable ILs. In addition, the entire MIM tunnel junction fabrication procedure must be *in situ* to minimize the formation of native oxides or other IL layers upon exposure to air. This demands not only integration of the ALD with UHV physical vapor deposition (PVD) systems, but also integration of such an ALD-UHV PVD system with various surface engineering and characterization approaches for *in situ* fabrication and characterization of MIM tunnel junctions with control of the physical and chemical properties at atomic to sub-nanometer scales.

It should be noted that such An ALD-UHV PVD system must address the challenges caused by the incompatibility between the processing conditions for ALD and UHV PVD. One of the challenges is in the difference in vacuum ranges required for ALD and UHV PVD. ALD is operated at low vacuum, which means the metal film deposition in UHV PVD and dielectric deposition in ALD must be carried out in separate chambers to prevent the contamination of sputtering sources with the ALD chemical sources. An UHV interface between these chambers is hence necessary to allow *in situ* transportation of large wafers. In addition, ALD involves active chemical vapors, which means the ALD gas delivery needs to be handled in a safe manner to prevent contamination of the UHV PVD chamber. A separate chamber becomes important in serving several purposes of 1) separation of the ALD and UHV PVD, 2) loading and unloading samples; and 3) *in situ* modification and characterization of the sample surfaces as well as interfaces. Another challenge is in different sample temperatures required for ALD and UHV PVD process. Taking Nb/Al/Al$_2$O$_3$/Nb SIS JJs as an example, the UHV magnetron sputtering of the Nb electrodes is carried out at temperatures below room temperature. A cooled sample stage is adopted to host the sample chuck and a specially designed sample chuck engagement is required to ensure an excellent thermal link between the stage and chuck. Hosting the same sample chuck in the ALD system is therefore preferable and uniform heating and ALD source distribution across such a sample assembly is a difficulty that needs to be overcome in the integrated ALD-UHV PLD system. Commercial systems that meet these requirements can be cost prohibitive to many research labs. Furthermore, the literature on the design of such an integrated ALD-UHV PVD system is, until now, nonexistent. In order to address the need in research and development of high-quality tunnel junctions, this paper describes an ALD-UHV PVD system that integrates a viscous-flow ALD module to an UHV sputtering chamber via a load lock (ALD-UHV sputtering in the rest of the paper) that allows for modification of the interfaces and *in situ* characterization. In fact, the interface designed on our ALD module is compatible with most HV and UHV chambers, allowing this ALD system to be integrated with many other PVD or CVD chambers.  As an illustration, fabrication of Nb/Al/Al$_2$O$_3$/Nb SIS JJs was carried out in this home-built ALD-UHV sputtering system, and the results taken on these devices are reported.

## II. ALD-UHV SPUTTERING SYSTEM DESCRIPTION

### A. System Overview

The ALD-UHV sputtering system developed in this work has four main components; a viscous flow ALD chamber, a UHV sputtering chamber, a load lock, and a sample transportation system. FIG 2 shows the layout of these components. In particular, the geometries of the sample at different stages of the *in situ* fabrication are depicted including UHV sputtering of metals (FIG 2a), possible surface/interface treatment with plasma in the load lock (FIG 2b), and ALD growth of the tunnel barrier (FIG 2c) with sample transfer from chamber to chamber provided by the sample transport system (FIG 2d). Three gate valves shown in FIG 2 allow each of the three cambers for sputtering, load lock, and ALD to be sealed completely during the corresponding operations. Not shown in FIG 2 is a second UHV sputtering/ion beam chamber for MTJs, which is connected to the left of the first UHV sputtering for superconductors (FIG 2a). With another sample transportation system attached to the second sputtering chamber, a sample can be transported between different chambers for fabrication of JJs, MTJs, and more complicated devices such as magnetic Josephson junctions [27,28]. This means the current design of the ALD-UHV sputtering system is very versatile and a cluster of UHV chambers may be integrated to this system for  fabrication of multiple functional materials.

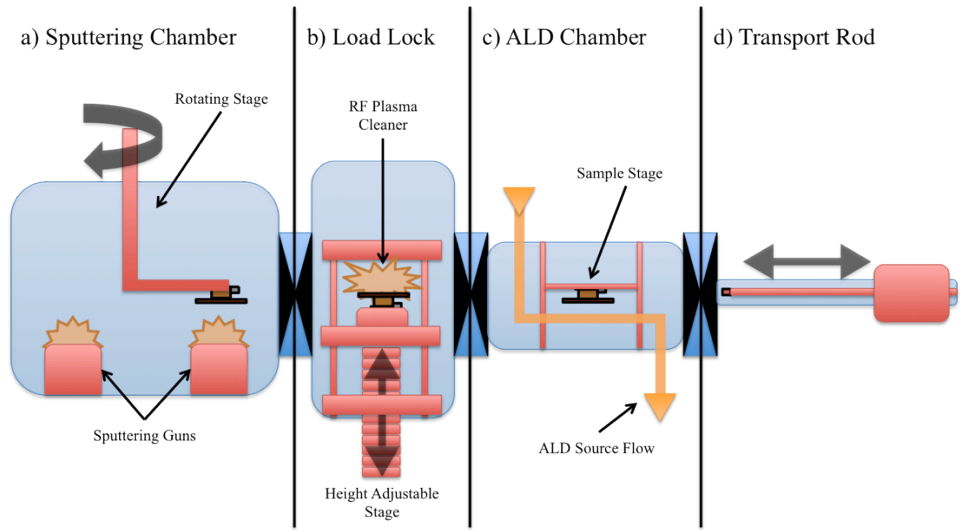

**FIG 2**: A schematic diagram of the ALD-UHV sputtering system. The UHV sputtering chamber (a) has three sputtering guns and a rotatable, water-cooled stage. It is kept at constant UHV by a cryopump. The load lock (b) is used for sample loading and unloading and contains an RF plasma treatment stage with an adjustable height stage. It is quickly brought between atmospheric pressure and high vacuum by a turbomolecular pump. The ALD chamber (c) is isolated from the other chambers by two gate valves to ensure a proper flow profile of the ALD sources, and it is heated by heat rope wrapped around its exterior (not shown). ALD is a low vacuum process, so the ALD chamber is pumped by a mechanical pump. Finally, the magnetically coupled transport rod (d) allows for UHV sample transfer from one chamber to another.

**B. ALD Chamber**

The ALD chamber is a viscous flow reactor with the ALD source handling similar to that previously described by Elam et al[29]. However, our ALD chamber differs significantly from those previous reported in its dimensions, sample mounting, and heating system. In order to accommodate 2 inch wafers and to integrate with the necessary UHV gate valves, the chamber itself is constructed from a 3 inch outer diameter, 8 inch long stainless steel tube with two CF flanges welded to the two ends for interfacing with the load lock and sample transport through gate valves. On the inner side of the tube, rails were installed to catch a specially designed sample chuck (described in detail in section II.E). During ALD growth, the rails suspend the entire chuck with the wafer at the center of the chamber. The chamber is blackbody heated, instead of the common solution of using a pancake heater to heat only the sample, by heat tape wrapped around the external wall of the chamber. This hot-wall ALD chamber has advantages in uniform sample heating across the wafer and much reduced condensation of the ALD sources on the chamber wall. This design can be readily expanded to accommodate larger wafers using larger stainless steel tubes to make the ALD chamber.

A schematic diagram of the chamber is shown in FIG 3. FIG 3a presents a cartoon schematic to demonstrate the layout of the relevant sensors, gate valves, and heaters. FIG 3b and FIG 3c show engineering schematics from isometric and front angles, respectively, with the main chamber tube removed to show the inner details of the chamber to scale. For completeness, the ALD valves are presented schematically in FIG 3d, and will be discussed presently. As seen in FIG 3a, on either side of the ALD chamber are two 4.625 inch CF flanged UHV gate valves which allow the ALD system to be totally isolated from the other chambers during operation. One of these gates is interfaced directly to the load lock (described in section II.D) while the other is interfaced to the transport rod (described in section II.E). These gate valves are critical to avoid contamination of the other chambers and components with ALD films and source chemicals. Sample docking is achieved by rails housed in a cylindrical inset which is held inside the ALD chamber with set screws (FIG 3b). There are four smaller flanges in the chamber. The top and bottom flange are for source delivery and exhaust, respectively. The other two small flanges are for a thermocouple and a quartz crystal monitor (QCM).

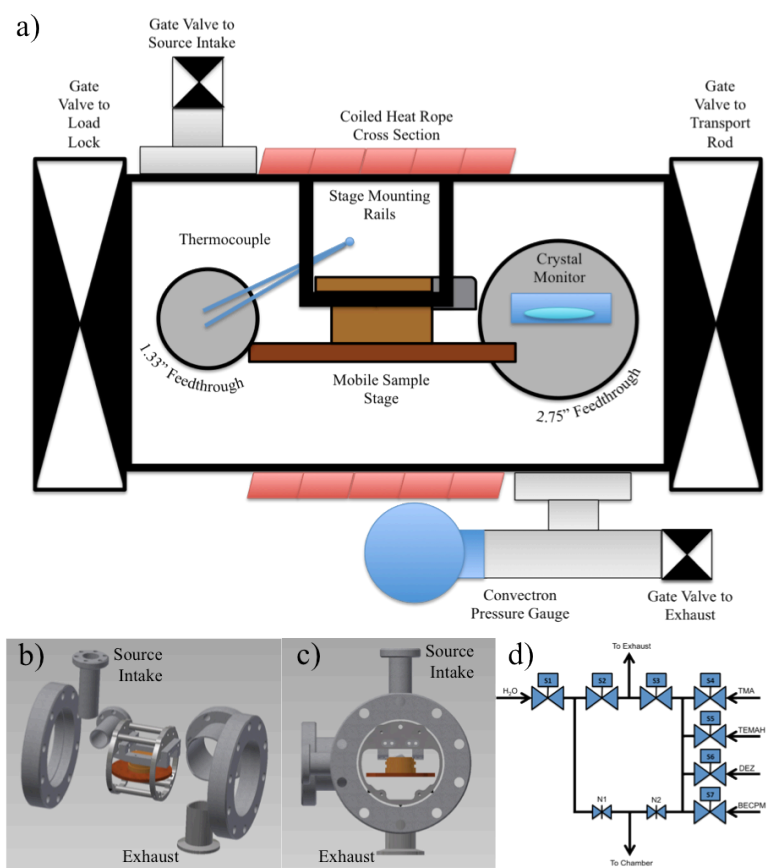

**FIG 3**: A schematic cross section (a) isometric (b) and front (c) views of the ALD chamber with the main chamber body removed. The sample stage (copper) is mounted on a cylindrical rail assembly. The sources are delivered through a computer controlled solenoid manifold (d) and a 1.33 inch CF flange on the top of the chamber. The internal temperature of the chamber is measured via thermocouple mounted to a 1.33 inch CF flange on the front left side of the chamber. Growth is monitored with a quartz crystal monitor (QCM) mounted on a 2.75 inch CF flange on the back left side of the chamber. The exhaust port is a QF flange on the bottom for the chamber, and it is fitted with a convectron pressure gauge. Gate vales to the load lock are mounted on 4.625 inch CF flanges on the front and back of the chamber.

ALD usually occurs at elevated temperatures, inside the chemical reaction window or "ALD window" defined by the given precursors [17]. To achieve these temperatures, the chamber is heated using resistive heat tape (McMaster-Carr) wrapped around the outside of the chamber. The power to the heater was provided by a variac and the temperature of the ALD chamber wall was controlled using an Omega temperature controller with feedback temperature reading from a K-type thermocouple inside the chamber. This blackbody heating is a simple and cost effective solution which replaces the more popular pancake heater in commercial ALD systems. The benefit of the blackbody heating is in flexibility of sample chuck design. In the specific case of Nb/Al/$Al_2O_3$/Nb SIS JJs, the chuck needs to be cooled efficiently during Nb electrode sputtering and can be readily heated to 200-300 ºC in the ALD chamber. An additional benefit of this blackbody heating for ALD, as we have mentioned earlier, is in ease of expansion of the chamber for larger wafers while maintain the uniformity of heating across the wafer. The source delivery tubing was also heated with resistive tape to prevent condensation of the sources. In order to achieve ALD growth, the delivery tubing must be heated to above the boiling point of the source but below its decomposition temperature. Heating the delivery tubing is particularly important to increase the flow of the sources with low vapor pressures such as TEMAH, which is used for ALD growth of $HfO_2$. An example of the temperature of the chamber and one section of tubing while preheating for ALD-$Al_2O_3$ growth is given in FIG 4. ALD $Al_2O_3$ growth occurs best around 200 ºC, which the chamber (top curve, black circles) achieved through blackbody heating after only 90 minutes of heating at 150 W. For the delivery tubing (bottom curve, red squares), a temperature of 90 ºC was selected for TMA to minimize condensation. Because of the low thermal mass of the tubing, this temperature can be achieved and adjusted quickly, as shown by the shoulder in the curve at 40 minutes when the delivered power was increased. Using heat tape and blackbody radiation to heat the ALD chamber is a cost effective strategy for achieving a uniform temperature for an arbitrary chamber size, and heat tape on the delivery tubing allows fast changes in temperature for the sequential deposition of ALD films using different sources.

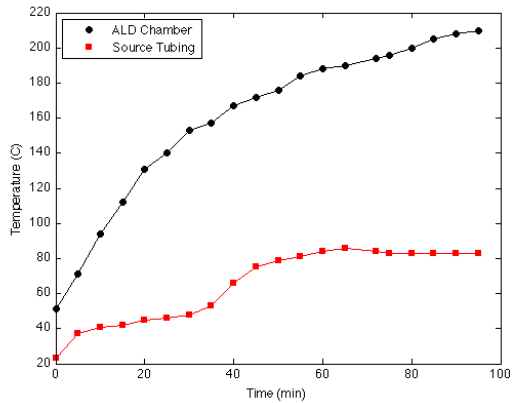

**FIG 4**: Temperature vs. time while heating the ALD chamber (top, black circles) and delivery line (bottom, red squares). The ALD chamber is heated externally with resistive heaters. The temperature is measured via internal thermocouple which is heated only by radiation from the chamber walls. The delivery tubing is heated in the same way, but measured by external thermocouple. Its smaller thermal mass allows faster, more responsive heating, as seen by the shoulder in the data when the delivered current was adjusted.

At sample temperatures within the ALD window, the reaction between the ALD sources produces self-limited growth, but only if the sources are delivered into the chamber correctly. A schematic of the valve system is given in FIG 3d. The delivery tubing is made from ¼" outer diameter (OD) seamless stainless steel tubing and stainless steel compression fittings (Swagelock). The sources enter into the delivery tubing through high speed solenoid valves (Parker Series 99) with a switching time of 100 ms, which are controlled by a custom solenoid controller and a labview program. The pressure pulse height of the sources in the chamber is controlled by two needle valves (Swagelok). Installed sources (FIG 3d S1, S4-S7) include $H_2O$ (Ultima grade, Fischer Scientific), TMA (Sigma-Aldrich) for $Al_2O_3$, TEMAH (Sigma-Alrich) for $HfO_2$, Diethyl Zinc (DEZ, Akzo Nobel) for ZnO, and $MgCp(Et)_2$ (BECPM, Strem Chemicals) for MgO. New sources can easily be added by simply installing another valve onto the already existing line of source valves, making this delivery system flexible and scalable. But during the ALD process, only one of these sources may enter the chamber at a time, and after such an exposure the chamber and delivery tubing must be purged. To do this, the source solenoids are closed and the exhaust solenoids (FIG 3d, S2 and S3) are opened. This creates a path from the valve assembly directly to exhaust to quickly purge the system of any remaining source vapor. Using ALD $AL_2O_3$ as an example, one ALD cycle consists of opening the $H_2O$ valve (S1), opening the purge valves (S2 and S3), opening the TMA valve (S4), followed by opening the purge valves again. Other films can be grown using an identical cycle, but replacing the TMA valve with another source's valve. Typically, 1-5 second source exposures and 30-60 second purges are performed. These are significantly higher than cycle times reported by other groups (typically reported values are 10s or 100s of milliseconds) due to the significantly longer tubing we used to satisfy safety guidelines. It is worth noting that these tubes should be as short as possible to minimize "dead volume" and to decrease the time it takes to complete one cycle. Further, any area that is exposed to both sources is in danger of becoming contaminated with the product of the bulk reaction between the sources, which in the case of water and TMA is a very fine alumina powder. So far, we have found no effective remedy for this problem aside from discarding contaminated parts.

To ensure the delivery system is operating correctly, a quartz crystal monitor (QCM) was installed inside the ALD chamber. QCMs are resonating quartz crystals with a resonant frequency that decreases when mass is added to the surface of the crystal, dampening its vibration. QCMs are often used to monitor the growth rate during PVD and CVD, but the resonant frequency of standard "AT cut" crystals is very sensitive to increases in temperature making them poorly suited for ALD growth at temperatures above 200 ºC. The quartz crystal used in this system is an "RC cut" crystal (Colnatech) specifically designed to withstand higher temperatures than "AT cut" crystals. FIG 5 shows the output from our QCM (FIG 5 top) and the pressure inside the chamber (FIG 5 bottom) as the sources are pulsed for 4 cycles. The first peak corresponds to a water exposure, and the exchange of $CH_3$ groups for OH groups, so the mass on the crystal changes very little, ~2 Hz/cycle. However, the second peak is a TMA dose, which corresponds to the deposition of Al, and as such the QCM frequency drops significantly, ~7 Hz/cycle. The sharp peaks in the QCM data are transient noise in response to the sudden change in chamber pressure. The QCM data shows a steadily decreasing linear trend at ~9 Hz/cycle, and the pressure pulses show consistent duration and magnitude. The QCM is therefore sensitive to sub-Angstrom changes in thickness and confirms that ALD growth is occurring consistently throughout the deposition.

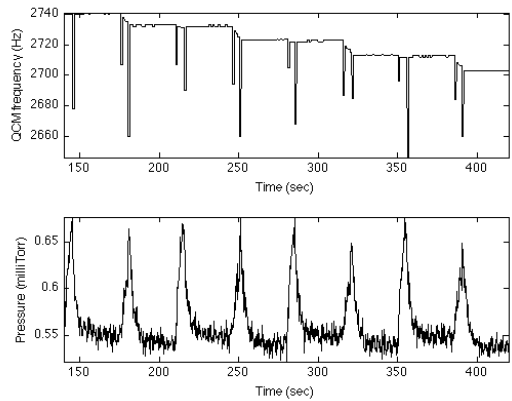

**FIG 5**: QCM frequency (top) and pressure data (bottom) of 4 ALD cycles. The QCM frequency drops ~9 Hz/cycle. The sharp spikes in the data are noise caused by a sudden increase in chamber pressure. The first pulse in the pressure data is H$_2$O, followed by TMA, and so on. The QCM frequency drops correspond exactly in time to the pressure pulses.

## C. Sputtering Chamber

On the other side of the load lock from the ALD chamber is the sputtering chamber. It is made from 20 inch OD stainless steel cylinder and is continuously kept at UHV, with a base pressure of ~$10^{-8}$ Torr or better with baking, by a cryopump (CTI cryogenics). The deposition pressure is controlled by a throttle valve (MKS type 653) and a mass flow controller (MKS type 1159B), and is typically 10-100 milliTorr. The pressure inside the chamber is sensed with a micro-ion gauge (Brooks Automation) for high vacuum up to $10^{-10}$ Torr, a convectron gauge (Brooks Automation) for low vacuum from atmospheric pressure to $10^{-4}$ Torr, and a capacitance manometer gauge (MKS Instruments) to accurately cover the sputtering pressure range. Multiple gas lines, including Ar, N$_2$, and O$_2$, enable sputtering of elemental sources, nitrides, and oxides. There are three magnetron sputtering guns in the chamber, two 3 inch guns and one 2 inch gun (Torus® from Kurt J. Lesker). The sputtering guns are mounted 90º apart and with their central axes 6 inches from the center of the chamber. They can be driven either with DC or RF power supplies. The DC power source (MDK 1.5K from Advanced Energy) has a useful range of 14 – 1500 W, while the RF power source (R601 from Kurt J. Lesker) has a useful range of 8 – 600 W at 13.56 MHz. The docking coupler in the sputtering chamber is a chilled-water cooled, copper U-shaped hard-stop which allows the sample chuck to enter and lock in place. The stage can be cooled to ~8 ºC, which is critical for depositing stress-free Nb films with good superconducting properties. The temperature is sensed with a k-type thermocouple (Omega Co). The stage, which rests ~6 cm above the sputtering target surface and on the central axis of the entire ALD-UHV sputtering system, can be manually moved into and out of the sputtering plasma with a rotating handle. This allows controlled exposure to individual sputtering guns. With three sputtering guns installed in the chamber, a wide variety of multilayer films may be grown including Al-wetted Nb, which is essential to Nb/Al/$Al_2O_3$/Nb JJs.

**D. Load Lock Chamber**

Integrating the UHV sputtering chamber with the ALD chamber was achieved through a load lock. The load lock, pictured in FIG 6a and FIG 6b, is made from an 8 inch OD stainless steel tubing with 10 inch CF flanges on its top and bottom, a 6 inch CF gate valve on its left that interfaces with the sputtering chamber, and a 4.265 inch CF gate valve that interfaces with the ALD chamber. It can be brought to ~$10^{-6}$ Torr in 10 minutes by a mechanically backed turbomolecular pump (TMP) from Edwards (nExt 240). This pressure can be held with an ion pump (Kurt J. Lesker LION 301) for vibration reduction. The pressures are sensed with a convectron gauge for low vacuum and a cold cathode gauge for high vacuum. An O-ring sealed door is installed on the front of the load lock for easy sample installation and removal.

Beyond its key role in connecting the ALD and sputtering chambers, the load lock has a docking chuck for RF plasma treatments, pictured in FIG 6b. The coupler chuck, pictured in FIG 6c, is made entirely of Teflon and has a U-shaped hard-stop identical to that in the sputtering chamber. A copper-beryllium spring rests inside the chuck to make electrical contact with the sample stage. A 20 gauge copper wire was driven through the Teflon to connect this spring with an RF electrical feedthrough to electrically isolate the sample stage from the chamber. Thus, during RF plasma treatments, the sample stage itself acts as one electrode. A removable, grounded stainless steel plate can rest above the docking chuck to act as the counter electrode. The distance between the sample stage and this steel plate can be controlled by a 6 inch stroke linear actuator which moves the chuck vertically (LSM from Kurt J. Lesker). The plasma is driven with an RF power supply and matching network (R601 from Kurt J. Lesker). Gas lines supply either $O_2$, $N_2$, or Ar for a variety of plasma treatment options including oxygen plasma cleaning and Ar ion milling. A metered gate valve on the TMP allows control of the load lock pressure. At 30 milliTorr of Argon, 150 W RF power, and a sample-to-electrode distance of 3 cm, an ion etch rate of 1 nm/min for Nb has been measured. An *in situ* plasma stage is indispensible to tunnel junction fabrication since hydrous plasma treatments are critical for achieving a quality interface when growing ALD films on noble metals, [26] and ion milling is often required to remove native oxides from metal films to make good electrical contact, i.e. Nb during JJ fabrication.

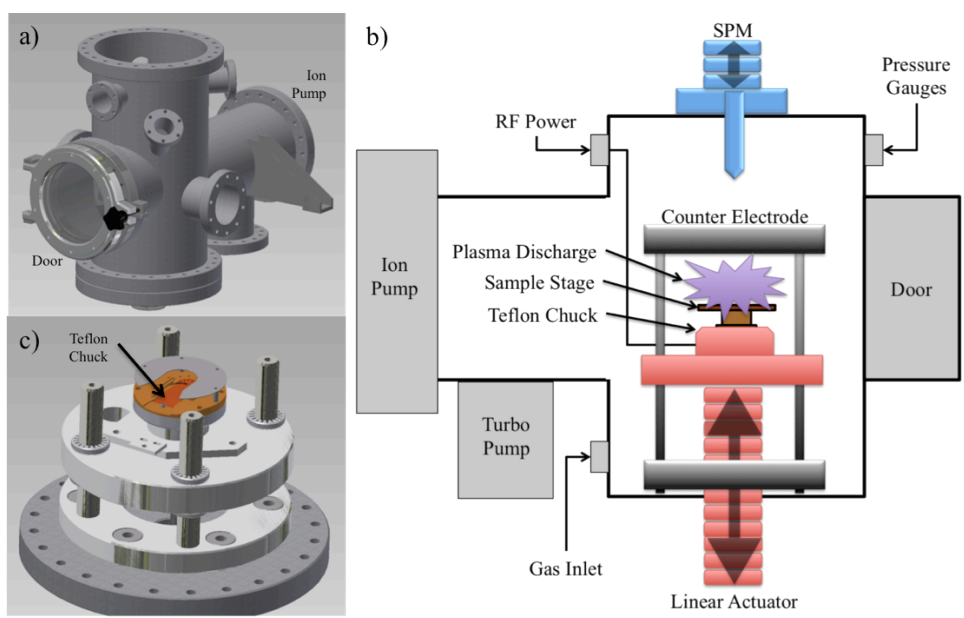

**FIG 6**: Engineering drawings of an external view of the load lock chamber (a), a schematic cross section of the load lock (b), and engineering drawings of the adjustable height chuck inside (c). The load lock (a, b) is crafted from an 8 inch OD stainless steel tube. The door on the front is sealed with an O-ring and allows for quick and easy transfer of the sample stage. The 10 inch flange on top is designed for scanning probe microscopy instrumentation such as atomic force microscopy or scanning tunneling microscopy. The rear flanges are for the installation of a turbomolecular pump for fast pumping and an ion pump for a vibration free vacuum. The sample chuck (c) is fitted with a Teflon, U-shaped hard stop so the sample stage can lock in place. Teflon was chosen to electrically isolate the sample from the chamber during RF plasma treatments. The chuck's height is adjustable so that the sample may be engaged by instrumentation, or to adjust processing parameters during a plasma treatment.

In addition to *in situ* plasma cleaning, this load lock was designed for *in situ* sample characterization with scanning probe microscopy (SPM), particularly RHK Technology's line of atomic force microscopes (AFM), the designed placement of which is pictured in FIG 6c. The U-shaped chuck in the load lock is compatible with RHK's sample chuck for easy sample transfer. The linear shift on this stage can raise the sample to the scan head, which could be mounted on the 10 inch CF flange on top of the load lock. Further, the ion pump in the chamber would allow for *in situ*, vibration free, high vacuum characterization of sputtered and ALD films at various points in their growth. Beyond AFM, the load lock could be configured for scanning tunneling microscopy or spectrographic ellipsometry, which would provide valuable insight into the microstructure, electrical, and optical properties of ALD films as they grow.

**E. Sample Transportation**

Seamless integration of the chambers, and transportation between them, is a critical function of this system. Each chamber is connected to its neighbors by a CF flanged circular gate valve. The ALD chamber uses the smallest gate valve on a

4.625 inch CF flange. The internal diameter of the valve is 3.52 inch, which sets the maximum sample stage diameter. However, this dimension could easily be increased by choosing a larger OD for the ALD chamber. Rectangular valves, or slot valves, are also available. But, the circular gate allows for a more versatile stage design, allowing easier integration with UHV-sputtering. The sample stage is moved through these gates, and from chamber to chamber, with a magnetically coupled linear shift (transport rod) from UHV Designs with a 3 feet stroke length.

The sample stage itself is shown in FIG 7 and has three main features of interest; the sample platform, the docking coupler, and the transportation coupler. The sample platform is a 2 inch copper disk with six equally spaced threaded holes on its perimeter which allow the sample to be clamped to the platform. Copper was chosen for its excellent thermal conductivity to create a uniform heat profile during ALD and water-chilled sputtering. The docking coupler is a copper, U-shaped protrusion screwed to the sample platform, and it acts as the male half of the stage docking mechanism. The U-shape allows for hard-stop docking in the sputtering chamber and load lock chucks, and it is compatible with RHK UHV scanning probe microscope systems. Slots are cut into the walls of this protrusion to allow soft-stop docking with the ALD chamber's rails. The stage transportation coupler is a stainless steel threaded hole that mates with the terminating screw on the transport rod (rod transportation coupler). This stage design, along with the hard-stop and soft-stop docking couplers in the sputtering chamber, load lock, and ALD chamber, allow for easy and reliable sample transfer from chamber to chamber.

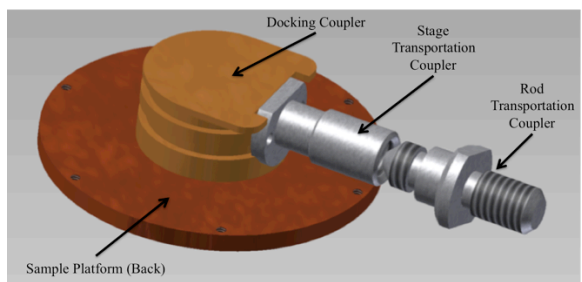

**FIG 7**: An engineering schematic of the sample stage. The sample is mounted onto the bottom side of a 2 inch disk of copper, chosen for its thermal conductivity. The docking coupler, also made of copper for its thermal conductivity, is screwed onto the 2 inch disk. The coupler's U-shape mates with the hard-stop chucks in the sputtering chamber and load lock, and it is compatible with RHK UHV scanning probe microscopy systems. Slots are cut into the side of the docking coupler to catch the mounting rails in the ALD chamber. The stage transportation coupler is stainless steel, and it is screwed to the back of the docking coupler. The rod transportation coupler is also stainless steel, and its dovetail shape helps to engage the sample stage.

## III. DEVICE FABRICATION

### A. Nb/Al/ALD-AL$_2$O$_3$/Nb Trilayer Fabrication

As a proof of concept of this ALD-UHV sputtering system, Nb/Al/ALD-AL$_2$O$_3$/Nb trilayers were fabricated and patterned into JJ arrays using advanced photolithography in combination with electron beam lithography. For comparisons, JJs made on traditional thermal Nb/Al/AlOx/Nb trilayers were also fabricated, in which Al is used as the source material for the thermally oxidized AlOx tunnel barrier. In the case of trilayers with ALD tunnel barriers, Al was also adopted as a wetting layer to facilitate the nucleation of ALD-Al$_2$O$_3$ and to prevent the formation of Niobium oxide (NbOx). NbOx has three oxidation states[30], only one of which is insulative[31,32], so preventing the formation of NbOx is critical to reproducible junction fabrication. The Nb films were sputtered in 14 mTorr Ar at 330 W DC, which yielded a power density of 46 W/in$^2$ and hence a high growth rate of 1.7 nm/s to minimize the formation of NbOx caused by traces of residual oxygen in the UHV sputtering chamber. For this work, the bottom electrode was 150 nm, and the top electrode was 50 nm. The Al wetting layer was sputtered in 14 mTorr Ar at 90 W DC to a thickness of 7 nm. The same thicknesses of Nb and Al were adopted in the thermal Nb/Al/AlOx/Nb trilayers. For the trilayer with an ALD tunnel barrier, 2 - 20 cycles (0.2 nm – 2.4 nm) of ALD-Al$_2$O$_3$ growth occurred at 200 ºC with TMA and H$_2$O. A reference trilayer was also made that went through the ALD heating and cooling cycle, but was not exposed to the ALD reaction. For the traditional trilayers with a thermal oxide tunnel barrier, the Al wetting layer was exposed to either 1 Torr or 100 Torr of O$_2$ for ~3.5 hours in the sputtering chamber before the top Nb was sputtered. These pressure-times correspond to target critical current densities of 500 A/cm$^2$ and 50 A/cm$^2$, respectively[4].

The surface morphology of these trilayers was studied with contact mode atomic force microscopy (AFM). The AFM characterizations show conformal growth of ALD-Al$_2$O$_3$ on top of the Al wetting layers. The surface of the bottom Nb layer is smooth with an average roughness R$_{rms}$ of ~1 nm. The Al wetting layers still have comparable smoothness with R$_{rms}$ of ~1.1 nm. With 14 cycles of ALD-Al$_2$O$_3$ the R$_{rms}$ is ~1.3 nm. These morphologies confirm the conformal development of the ALD films which easily copy the smoothness of the base surface.

CIPT measurements [33] were taken on the unpatterened trilayers to confirm the integrity of the tunnel barrier at room temperature. CIPT measurements were performed on trilayers with the number of ALD cycles ranging from 2-20. The reference trilayer with 0 cycles was also measured. In the latter case, the tunnel resistance was too low to measure using CIPT, indicating the heating/cooling process in ALD did not cause significant oxidation of the Al wetting layer. For the other trilayers, the tunneling resistance was clearly identified by CIPT. In fact, a monotonic increase of the tunneling resistance with the number of the ALD cycles has been observed[34]. In addition, uniform tunneling resistance with a small standard deviation of less than 10% was observed on most samples with diameters up to 50 mm confirming good control of the tunnel resistance by varying the number of ALD cycles.

## B. ALD Interfacial Layer Characterization

ALD-AL$_2$O$_3$ was grown on sputtered Al substrates to probe the nucleation and measure the thickness of any IL that may form during ALD growth. Two sets of samples were fabricated. For the first set, ~50 nm Al was sputtered in 14 mTorr Ar at 90 W DC, and 0 – 100 cycles of ALD-Al$_2$O$_3$ were grown. For the second set, 0.1 – 1.0 nm Al was sputtered in 14 mTorr Ar at 15 W DC, and 60 cycles of ALD-Al$_2$O$_3$ were grown. The ALD-Al$_2$O$_3$ films' morphologies was characterized with AFM, and their thicknesses were measured with spectroscopic ellipsometry.

FIG **8** presents *ex situ* AFM deflection images of the morphology of the native oxide on 50 nm sputtered Al (a and b) and 20 cycles of ALD-AL$_2$O$_3$ grown on 50 nm Al sputtered *in situ* (c and d). Surface roughness measurements yield $R_{rms}$ = 1.1 nm for the native oxide and $R_{rms}$ = 1.3 nm for the ALD film. Surface roughness measurements on 20-100 cycles of ALD-AL$_2$O$_3$ grown on Al all showed comparable $R_{rms}$ ~ 1 nm. These comparable roughness values between the Al native oxide and varied thicknesses of ALD-Al$_2$O$_3$ grown on Al confirm the highly conformal nature of ALD-Al$_2$O$_3$ grown on Al.

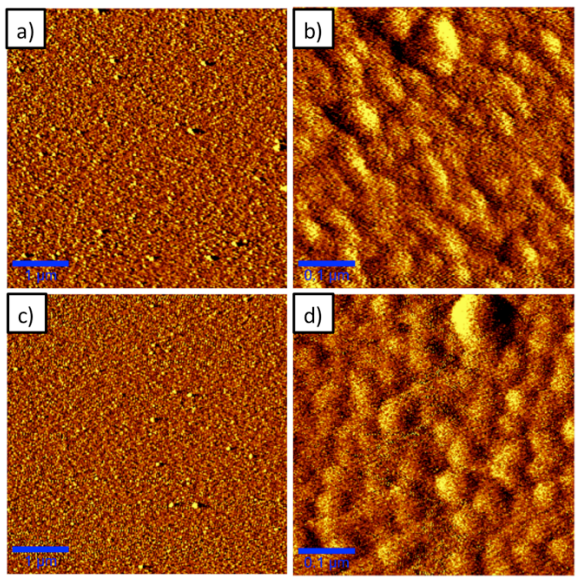

FIG 8: AFM deflection images of the native oxide on 50 nm sputtered Al (a and b) and 20 cycles of ALD Al$_2$O$_3$ grown on 50 nm *in situ* sputtered Al (c and d) over 5x5 μm (a and c) and 500x500 nm (b and d) scan windows. Topography measurements yield RMS roughness of 1.1 nm on the native oxide and 1.3 nm on the ALD oxide, confirming conformality. Scale bars: 1 μm (a, c); 0.1 μm (b, d).

The ellipsometry results from the ALD-Al$_2$O$_3$ grown on 50 nm Al, given in FIG **9** and expanded upon in an earlier paper [22], show a growth rate of 1.19 Å/cycle, which is consistent with previous reports of ALD Al$_2$O$_3$ on a variety of other substrates [17]. The non-zero y-intercept of the trendline indicates an IL of ~2 nm was formed. We hypothesize that this IL is thermally grown aluminum oxide (AlOx) that results from a combination of two growth mechanisms. The first mechanism is simple thermal oxidation while heating up the ALD chamber. In order to confirm this, a sample was heated to 200 ºC in ALD chamber over the heating and cooling course of ~1-2 hours without ALD. JJ arrays were fabricated using the same micro-fabrication procedure and the current-voltage characteristics of these reference JJs produced with 0 ALD cycles (presently discussed at length in III.C). The measured critical current density of 9.5 kA/cm$^2$ is far too large for a ~2 nm tunnel barrier, suggesting the formation of a substantial IL oxide during ALD heating is unlikely. The second mechanism we propose is diffusion oxidation during the ALD process. Since a bare Al surface is exposed to H$_2$O at 200 ºC, some oxidation will occur. The total thickness of this oxide depends on the temperature, the partial pressure of oxygen, and the total number of ALD cycles performed. These two mechanisms, in combination with oxidation upon exposure to ambient atmosphere, produce four scenarios for ultrathin ALD film growth on Al. FIG 10 is a cartoon representation of these scenarios. FIG 10a, shows an AlOx IL formed from *in situ* oxidation that occurs during the ALD heating process from traces of H$_2$O in a heated chamber; a very thin oxide is formed on the Al surface. If an ALD film is grown that is too thin to prevent diffusion of ambient oxygen, then a native oxide will form from *ex situ* oxidation underneath the ALD film when the sample is removed (FIG 10b). For longer ALD depositions and thicker films, ALD growth and diffusion oxidation will occur together during the ALD process, producing co-growth and a substaintial AlOx IL (FIG 10c). We hypothesize that the additional ~2 nm of Al$_2$O$_3$ in FIG **9** can be explained by co-growth, and this mechanism is consistent with previous reports of high growth rates during the nucleation of ALD on other, easily oxidized metals [23,35]. However, if a thin ALD film is grown and then capped with a diffusion barrier (such as an Al or Nb top electrode), both co-growth and ambient oxidation can be minimized, producing a tunnel barrier that is dominated by ALD growth. Co-growth is controlled by temperature, oxygen partial pressure, and the oxidizability of the substrate, so careful tuning of the ALD processing parameters or clever substrate engineering is therefore necessary to produce a crisp MI interface, even with *in situ* ALD.

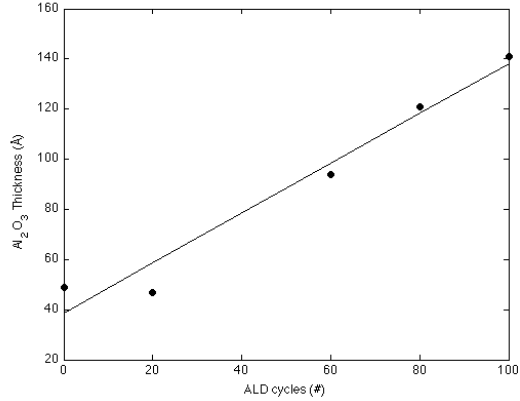

**FIG 9**: *Ex situ* ellipsometry of ALD Al$_2$O$_3$ on 50 nm in situ sputtered Al. From the slope of the trendline, the growth rate is 1.19 Å/cycle. The non-zero y-intercept at 23 Å indicates the presence of a thermally oxidized interfacial layer.

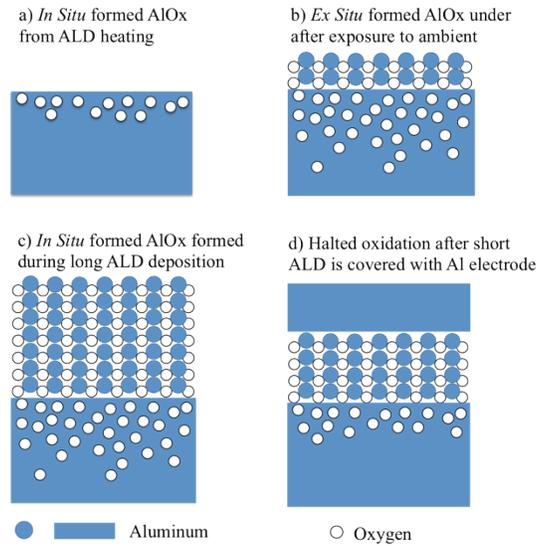

**FIG 10**: Four distinct oxidation regimes exist when growing ALD Al$_2$O$_3$ *in situ* on Al substrates. A very thin thermal oxide forms during the ALD heating process from trace H$_2$O in the chamber (a). If the ALD film is not thick enough to act as a diffusion barrier against ambient oxygen, a native oxide will form underneath the ALD film (b). During long ALD depositions, ALD growth and thermal oxidation occur together, which forms a significant interfacial layer (c). However, if a thin ALD film is capped with a diffusion barrier, both co-growth and ambient oxidation can be minimized (d), as is the case with tunnel junction fabrication.

To confirm these hypotheses, and to set a lower limit on the IL formed during *in situ* ALD, 60 cycles of ALD-Al$_2$O$_3$ was grown on ultrathin 0.15 – 1.0 nm Al, which was sputtered onto Si(100). The results of this study are given in FIG 11 and show a monotonic increase in ALD-Al$_2$O$_3$ thickness with increasing Al wetting layer thickness. This confirms that the IL is formed by the thermal oxidation of the underlying Al wetting layer. Further, by using only ~0.15 nm of Al as a wetting layer, the AlOx IL can be reduced to ~ 0.25 nm. Therefore, by using ultrathin Al wetting layers in JJs, the tunnel barrier thickness can be dominated by ALD-Al$_2$O$_3$ instead of the thermally oxidized IL.

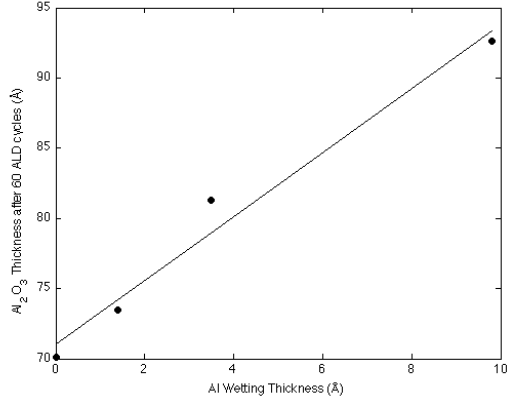

**FIG 11**: Ex Situ ellipsometry of 60 cycles ALD-Al2O3 on in situ sputtered 0 – 1 nm Al wetting layers sputtered onto Si(100). As the Al film thickness decreases, so does the total ALD-Al$_2$O$_3$ thickness, even though the same number of ALD cycles was performed on all samples. This indicates that the interfacial layer formed during ALD growth on Al is caused by the thermal oxidation of the underlying Al layer.

## C. Josephson Junction Fabrication and Characterization

To characterize the electrical performance of ALD-Al$_2$O$_3$ tunnel barriers, JJs were fabricated from Nb-Al/ALD-Al$_2$O$_3$/Nb trilayers, and JJs made from thermally oxidized Nb-Al/AlOx/Nb trilayers served as a reference. The number of ALD cycles was ranged from 5-13, and the thermally oxidized target J$_c$ was varied from 50 A/cm$^2$ to 500 A/cm$^2$. The nominal dimensions of the JJs ranged from 3x3 μm$^2$ to 10x10 μm$^2$, though 3D profilometry (Tencor P16) indicates that processing reduces these nominal dimensions by ~1.5 μm on each side. The resistance of JJs with nominal dimensions ranging from 7x7 μm$^2$ to 10x10 μm$^2$ was measured at room temperature using a 4 point probe station. FIG 12 shows a 10x optical micrograph of the test circuit used in the 4 point measurements. There is a small ( < 20 Ω ) residual resistance at room temperature in this 4 point configuration, and these residual resistances were measured directly on a sample that did not go through the junction definition processing. TABLE I shows the results of this room temperature analysis. According to the well-known Ambegaokar-Baratoff formula, $R_N = \pi\Delta / 2eJ_C A$, where $R_N$ is the normal state resistance of the JJ, $\Delta$ is the superconducting gap energy, e is the charge of an electron, $J_C$ is the critical current density of the junction, and A is the area of the junction. By fitting $R_N$ vs. 1/A, $J_C$ can be estimated. We see a monotonic decrease in Jc with increasing ALD thickness, from $J_C$ = 770 A/cm$^2$ for 5 cycles to $J_C$ = 32 A/cm$^2$ for 13 cycles. Further, the specific tunnel resistance ($R_N$A) is comparable across the range of areas for all tested JJs, and $R_NA$ increases monotonically with increasing ALD thickness. These results confirm that ALD produces a uniform tunnel barrier with sub-nanometer thickness control.

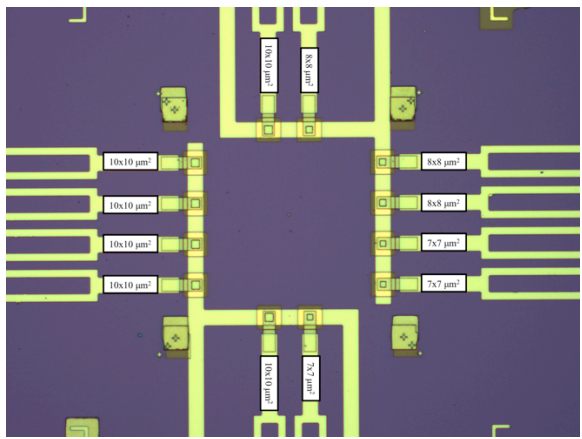

**FIG 12**: A 10x optical micrograph of the completed JJ chip. Each chip contains 12 JJs, which are either 10x10 μm², 8x8 μm², or 7x7 μm² nominal size. The bifurcated electrical leads allow for 4 point measurement with small residual resistances of < 20 Ω.

**TABLE I**: Estimated critical current density ($J_C$) and tunnel specific resistance ($R_NA$) of ALD JJs

| Tunnel Barrier | Minimum Barrier Thickness (Å) | Estimated $J_C$ (A/cm²) | $R_NA$ (μm Ω) |
|---|---|---|---|
| 5 cycles ALD $Al_2O_3$ | 6 | 696 ± 150 | 375 ± 66 |
| 8 cycles ALD $Al_2O_3$ | 9.6 | 371 ± 113 | 650 ± 150 |
| 10 cycles ALD $Al_2O_3$ | 1.2 | 38 ± 3.4 | 6750 ± 571 |
| 13 cycles ALD $Al_2O_3$ | 15.6 | 32 ± 2.7 | 11000 ± 1169 |

While room temperature characterizations of microstructure and resistivity provide important information about the quality of the ultrathin $Al_2O_3$ tunnel barrier grown by ALD, and the results indicate strongly the formation of a uniform, low-leakage tunneling barrier [34], a low temperature (below critical temperature of the superconducting electrodes) measurement of quasi-particle tunneling characteristics is the ultimate test to determine the integrity of the tunnel barrier. In order to measure the quasi-particle tunneling spectra, SIS tunnel junctions were fabricated from a trilayer with an 8-cycle ALD barrier layer on a 7 nm Al wetting layer (trilayer A) and the 0-cycle reference sample (trilayer B) which only went through the heating and cooling steps of ALD and also had a 7 nm Al wetting layer. Square junctions with nominal dimensions ranging from 3x3 μm² to 5x5 μm² were made and tested using a low noise SIS tunnel junction measurement system [36]. FIG 13 shows the IVC of three Nb/Al/ALD-$Al_2O_3$/Nb junctions with varying dimensions. The low subgap leakage current and uniform specific tunnel resistance $R_NA$ = 3.57 kΩ · μm² at voltages greater than 2Δ/e, where Δ and e are the superconducting gap energy of Nb and the charge on an electron, verify that eight cycles of ALD $Al_2O_3$ formed a uniform, low-leakage tunnel barrier. Despite the low subgap leakage current and the uniform specific tunnel resistance of both trilayers, the expected supercurrent at zero voltage due to the Josephson effect is entirely suppressed on trilayer A and heavily suppressed on trilayer B. The magnitude of this supercurrent, $I_c$, is expected to be ~75% of the gap current, $I_g$, defined as the current at the gap voltage of $2\Delta/e$ [37]. For trilayer B, $I_c$ is only ~30% $I_g$; for trilayer A, $I_c$ is nonexistent. The tunnel barrier for trilayer A terminates in a hydroxylated surface due to the chemistry of ALD, and residual water in the ALD chamber almost certainly hydroxylated trilayer B during the heating process. We speculate that these hydroxyl groups act as charged scattering centers for Cooper pairs, and this is the source of the apparent pair breaking mechanism across these tunnel junctions.

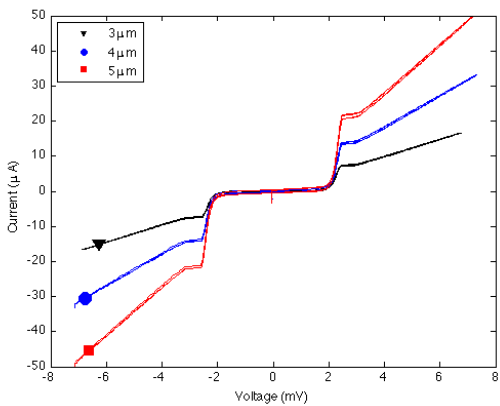

**FIG 13**: A low temperature current-voltage curve for three Josephson junctions with tunnel barriers fabricated by 8 cycles of atomic layer deposition (trilayer A). The square junctions were fabricated with sides of 3um (bottom, red square), 4um (middle, blue circle), and 5um (top, black triangle). The current densities for these junctions are identical, indicating a uniform tunneling barrier has been grown. The cooper pair tunneling current, which should be seen at 0 voltage, is absent despite significant single electron tunneling. This indicates an unknown mechanism is suppressing cooper pair tunneling.

## IV. CONCLUSION

To conclude, producing a ~1 nm leak-free tunnel barrier on metallic electrodes with a pristine interface is an outstanding challenge in the development of advanced electronics including JJs, MTJs, MJJs, and other devices. To address this challenge, a home-designed viscous-flow ALD module has been interfaced with an UHV sputtering chamber for *in situ* fabrication of MIM multilayer stacks. A sample transportation system, including a linear transport rod, a load lock, and an SPM compatible sample stage were developed and implemented. Using this ALD-UHV sputtering system, we have investigated the suitability of using Al as a wetting layer in SIS Nb-Al/$Al_2O_3$/Nb JJs, and found that while a thermally oxidized interfacial layer may form, it can be minimized by reducing the Al wetting thickness to 0.15 nm for ~1 nm thick $Al_2O_3$ tunnel barrier. Further,

SIS Nb-Al/Al$_2$O$_3$/Nb JJs were fabricated and characterized. We have shown that uniform, leak-free tunnel barriers and the critical current density and specific tunnel resistance can be controlled by altering the thickness of the ALD tunnel barrier in the range of 0.6 nm to 1.6 nm. This result demonstrates the viability of this integrated ALD-UHV sputtering system for the fabrication of tunneling devices comprised of metal-dielectric-metal trilayers and multilayers.

## V. ACKNOWLEDGEMENTS

The authors acknowledge support in part by NASA contract NNX13AD42A, ARO contract No. ARO-W911NF-12-1-0412, and NSF contracts Nos. NSF-DMR-1105986 and NSF EPSCoR-0903806, and matching support from the State of Kansas through Kansas Technology Enterprise Corporation. We would especially like to thank Physics Machine Shop machinist and research technologist, Allen Hase, for his tremendous help with fabrication and design.